\documentclass
[aps,floats,showpacs,preprintnumbers,amssymb,floatfix,nofootinbib,balancelastpage,superscriptaddress,amsmath]{revtex4}

\usepackage{epsfig,epstopdf}
\begin{document}

\preprint{IRFU-10-285}

\title{Seeking particle dark matter in the TeV sky}


\author{Pierre Brun}
\address{CEA/Irfu, Centre de Saclay, F-91191 Gif-sur-Yvette, France}

\begin{abstract}
Under the assumption that dark matter is made of new particles, annihilations of those are required to reproduce the correct dark
matter abundance in the Universe. This process can occur in dense regions of our Galaxy such as the Galactic center, dwarf galaxies
and other types of sub-haloes. High-energy gamma-rays are expected to be produced in dark matter particle collisions and could be detected by ground-based Cherenkov telescopes such as HESS, MAGIC and VERITAS. The main experimental challenges to get
constraints on particle dark matter models are reviewed, making explicit the pros and cons that are inherent to this technique, together with
the current results from running observatories. Main results concerning dark matter searches towards selected targets with Cherenkov telescopes are presented. Eventually, a focus is made on a new way to perform a search for Galactic sub-halos with such telescopes, based on wide-field surveys, as well as future prospects.

\end{abstract}

\maketitle


\section{Introduction}

Both from the ground and using satellites, observations of the cosmos in $\gamma$-rays have known significant improvement in the last twenty years. The Cherenkov technique has now reached a mature state, with several running observatories around the globe. The very good sensitivities of these instruments allow not only to observe $\gamma$-ray sources such as supernovae remnants, pulsars or active galactic nuclei, but also to search for exotic signals, in particular dark matter induced. In this paper, an attempt is made to describe the minimal knowledge that is necessary to have about this experimental technique to understand its pros and cons in the quest for non baryonic dark matter.

In the first section, the technique of Cherenkov telescopes is described, with an emphasis on the different methods to reduce backgrounds.
The second section is devoted to the expectations for possible dark matter signals, the way to choose targets and to display the constraints. The third and fourth sections deal with a review of results from all experiments using targeted searches and blind searches, as well as some prospects.

\section{Gamma-ray astronomy with Cherenkov telescopes}

\subsection{Principle}

\begin{figure}[h]
\includegraphics[width=.36\textwidth]{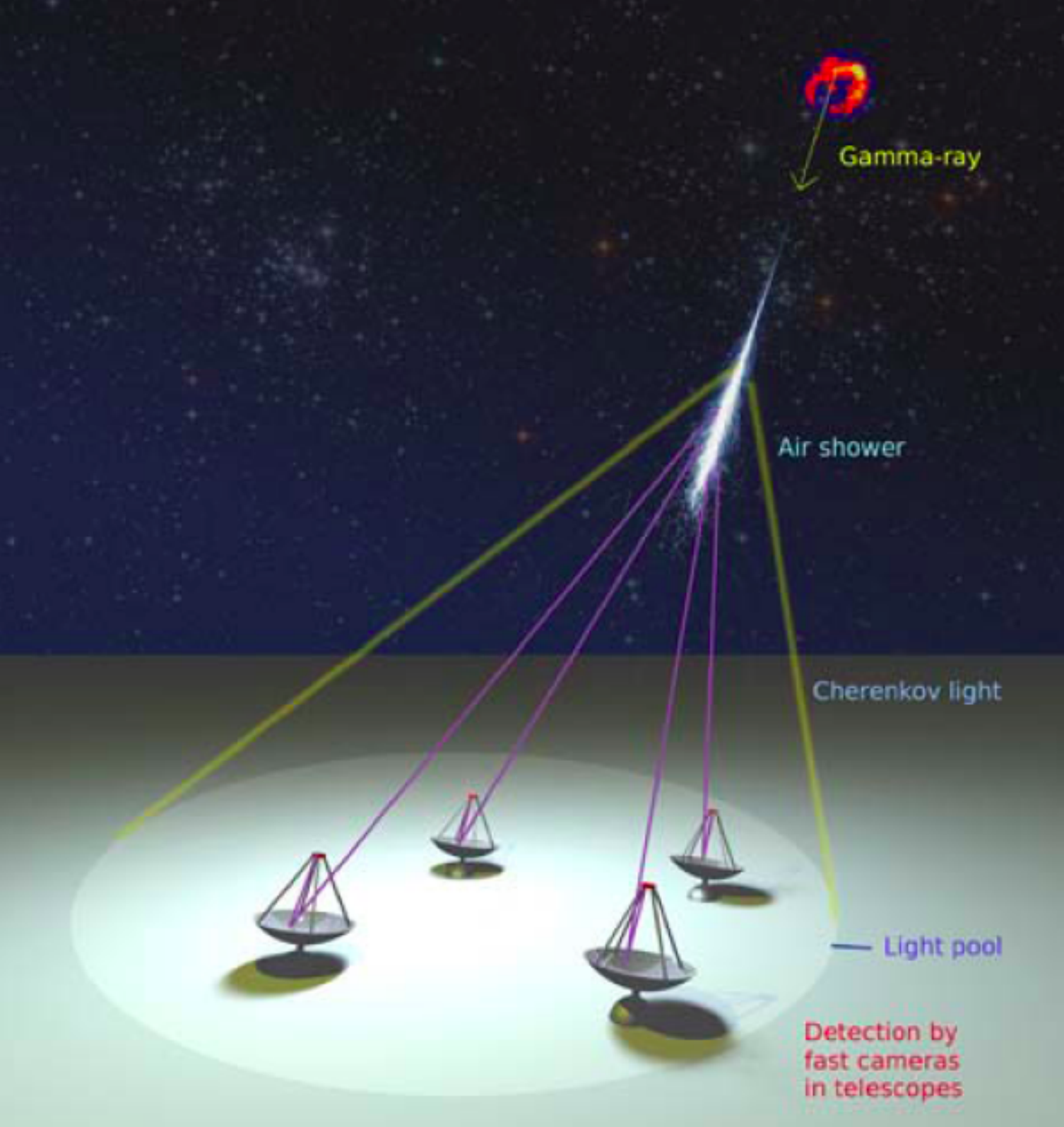}
 \includegraphics[width=.63\textwidth]{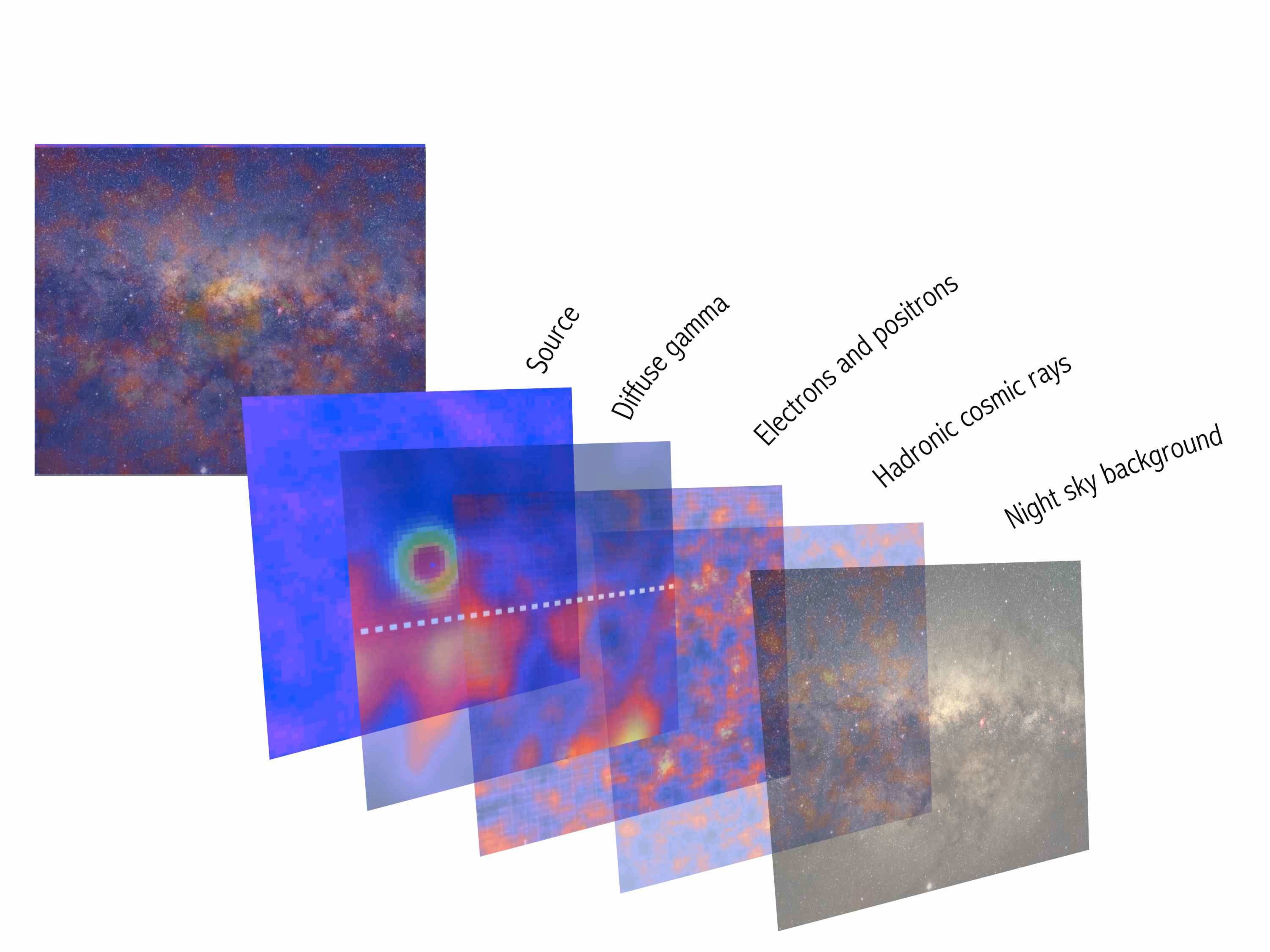}
  \caption{Left: Principle of the Cherenkov telescopes, several telescopes measure the Cherenkov flash induced by the initial $\gamma$-ray (image from K. Bern\"ohr~\cite{springerlink:10.1007/s10686-009-9151-z}). Right: Schematic representation of the different layers of background that Cherenkov astronomers have to suppress to obtain a image of the TeV sources. \label{principle}}

\end{figure}

The basic idea of running ground based telescopes to observe cosmic $\gamma$-rays is to use the atmosphere as a calorimeter. When a high energy particle hits the top of the atmosphere, it induces a cascade of secondary particles. At energies of about 1 TeV, that cascade is fully contained in the atmosphere, and it produces a flash of Cherenkov photons. Hadronic and electromagnetic particles produce different types of cascades. While hadrons induce irregular particle showers, electrons, positrons and $\gamma$-rays produce a more even shower. The Cherenkov flash is contained in a cone of 1$^\circ$ opening angle produced at 10 km height. The projection of the cone on the ground is a disk of order 250 m diameter, and the flash lasts about 5 ns. From any place inside this disk, the atmospheric shower is observable, should one use a sensitive enough instrument. Ground-based $\gamma$-ray observatories use this principle to measure $\gamma$-ray induced Cherenkov light, as sketched on the left panel of Fig.~\ref{principle}. Large dishes are used to collect enough photons, those are focused on very sensitive cameras equipped with photomultipliers. The cameras are able to integrate the signal very quickly and to resolve the image of the atmospheric showers, in order to fight against different types of backgrounds, as we shall see in the following. To gain in angular resolution, energy resolution and background subtraction, several telescopes are used simultaneously to observe the event, thus getting a stereoscopic view of the particle cascade.

\subsection{Backgrounds, data analysis and caveats of the technique}

As it can be inferred from the description of the technique in the previous subsection, the main backgrounds one has to fight against in order to have an image of a TeV source in the sky are:

\begin{itemize}
\item The night-sky background : diffuse light from stars, light pollution, the Moon, etc
\item Cherenkov light induced by hadronic cosmic rays
\item Cherenkov light from electron and positrons cosmic rays
\item Diffuse $\gamma$-ray emission that could overcome the signal from the source
\end{itemize}

Schematically, the image of the source is obtained when all these backgrounds are mastered, as shown on the sketch of Fig.~\ref{principle}. We shall now describe the basic principles that allow to suppress these backgrounds (see~\cite{Aharonian:2006pe} for a pedagogical application).

\begin{figure}[h]
  \includegraphics[height=.23\textheight]{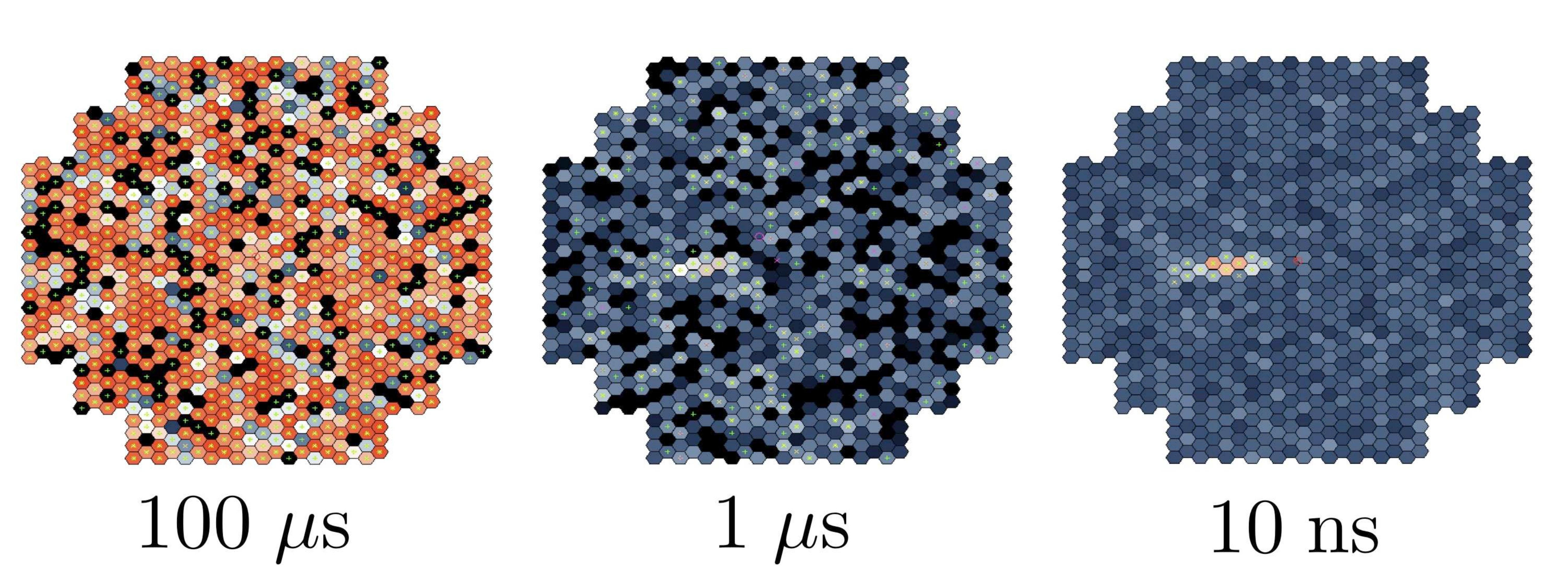}
  \caption{Three simulated image of a $\gamma$-ray shower, obtained with different integration times: from left to right 100$\mu$s, 1$\mu$s and 10 ns (courtesy K. Bernl\"ohr).\label{integration}}
\end{figure}

The night sky background (NSB) is fought against mainly using fast integration electronics. The closer the observation to the 5 ns of the actual shower flash, the less NSB is integrated. Fig.~\ref{integration} shows simulations of images of an atmospheric event obtained in the camera of a Cherenkov telescope. Here the same simulated $\gamma$-ray induced signal is integrated over 100 $\mu$s, 1 $\mu$s and 10 ns. For the longest integration times, the NSB signal dominates, whereas the electromagnetic shower appears clearly in the camera when the integration is done fast.
Depending on the experiments, different techniques are used (ADC flashes, analog memories), and the actual integration of the signal is of order 10 ns at best. Another way to suppress NSB is to use a triggering strategy that favors signals that are clustered, as the NSB is expected to be fairly constant over the field of view of the cameras.

\begin{figure}[h]
  \includegraphics[height=.24\textheight]{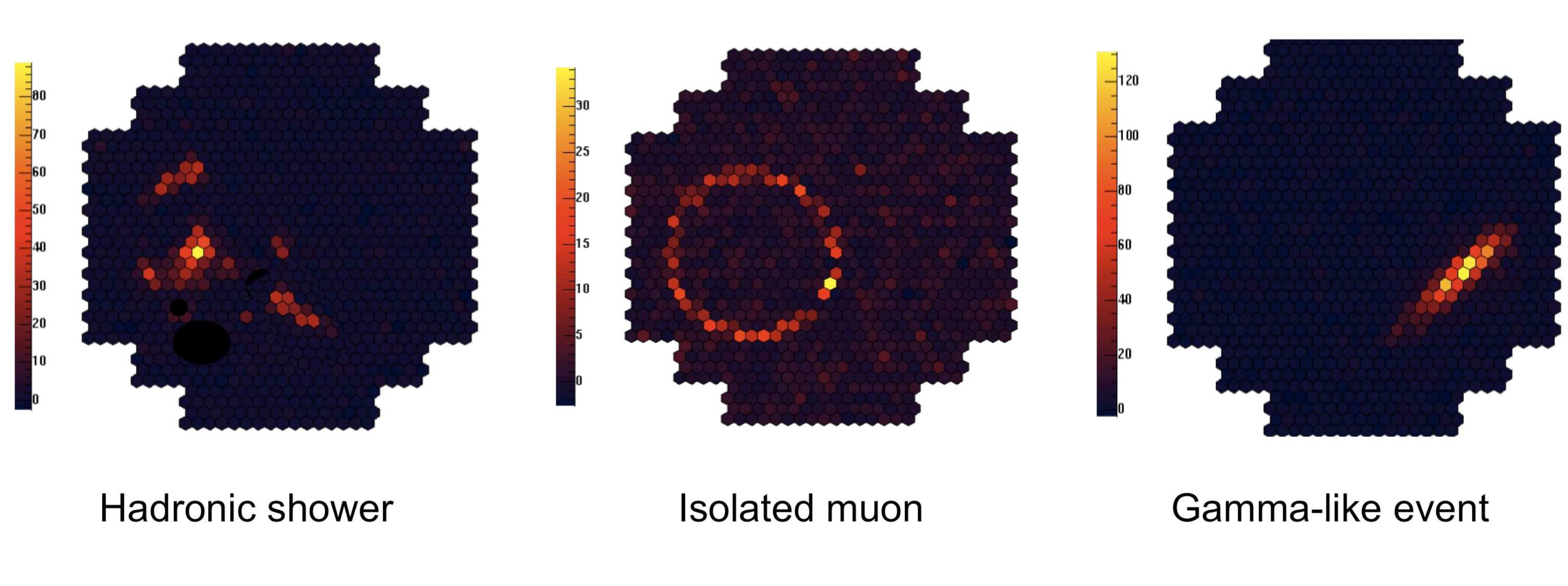}  
  \caption{Simulated images of atmospheric events induced from cosmic particles, as observed in the focal plane of a Cherenkov telescope. From left to right: hadronic shower, isolated muon, $\gamma$-like event (courtesy J. Hinton). \label{images}}
\end{figure}

Other types of background are mainly suppressed during the offline analysis. First, one has to get rid of the images of cosmic ray induced showers. This is handled by the capability of the telescopes to imaging the atmospheric event. Because heavy cosmic rays induce irregular showers, their images can be rejected during data analysis. In Fig.~\ref{images}, different images obtained in the focal plane of a telescope are displayed. In this figure, the left panel shows an image of a hadronic shower, the central panel is the image of a single muon ring (from a hadronic shower that is outside the telescope field of view), and the right panel is the image of a $\gamma$-like event. The offline analysis tools are designed to reject the two first ones, and keep the last one, based on the topology of the image. To get an idea of the rejection power of the analysis, for a typical bright source, 10 h of observation yield $10^7$ recorded events, out of which $10^4$ are actual $\gamma$-rays. After the topology-based analysis procedure, $10^5$ events are selected, dubbed $\gamma$-like events. At this point of the analysis, electrons and positrons cosmic rays are still present in the sample.

\begin{figure}
  \includegraphics[height=.4\textheight]{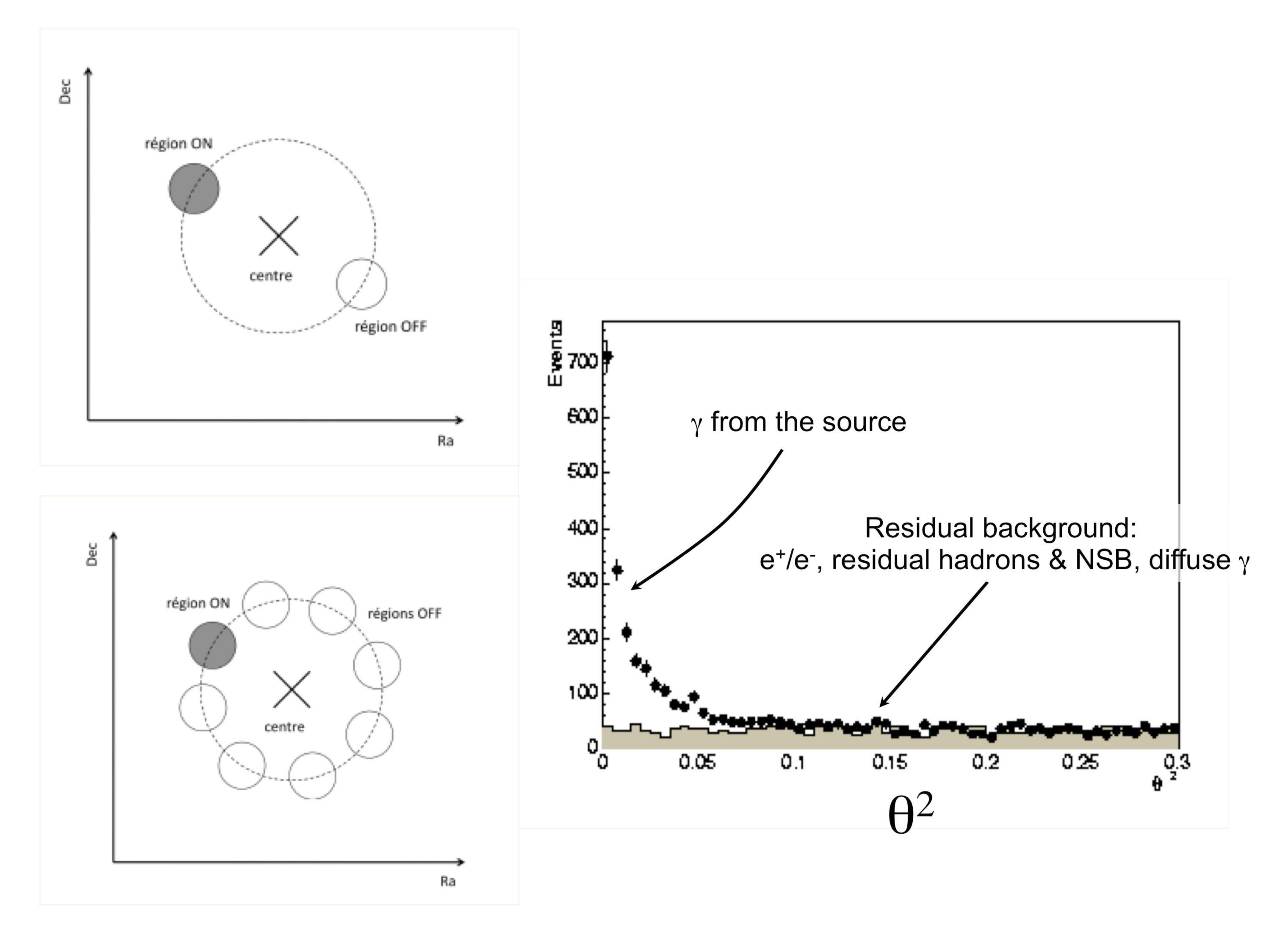}
  \caption{Illustration of the ON-OFF background suppression technique (figure from~\cite{vivier}).\label{onoff}}
\end{figure}

At fixed energy, electrons, positrons and $\gamma$-rays all induce identical electromagnetic showers, thus $\gamma$-like events contain all these species. A feature of charged TeV scale cosmic rays is that their distribution at the top of the atmosphere is isotropic. The subtraction technique is then based on the estimate of the isotropic part in a  region of the sky where no $\gamma$-ray signal is expected. This region is called the OFF region, and the corresponding event rate is subtracted where the source is expected to be (ON region). Fig.~\ref{onoff} displays two possible choices for OFF regions, and a histogram showing the signal from the source and the background. The later is composed of electrons, positrons, residuals from misidentified hadronic showers. Note that if an actual $\gamma$-ray signal is diffuse all over the field of view, this technique will likely erase it. The general rule is that a diffuse signals is observable with Cherenkov telescopes provided its extension is smaller than the field of view of the experiment, which is about a few degrees. Observing larger scales diffuse emission requires non standard techniques such as dedicated OFF runs.

The energy of the initial $\gamma$-ray is deduced from the amount of collected Cherenkov photons, and comparison to Monte Carlo simulations matching the conditions of data taking. Uncertainties lie in the knowledge of the atmospheric conditions, simulation uncertainties and intrinsic shower fluctuations. At the end, the energy resolution 
is about 10\%.

\subsection{Ongoing experiments}

\begin{figure}
  \includegraphics[height=.35\textheight]{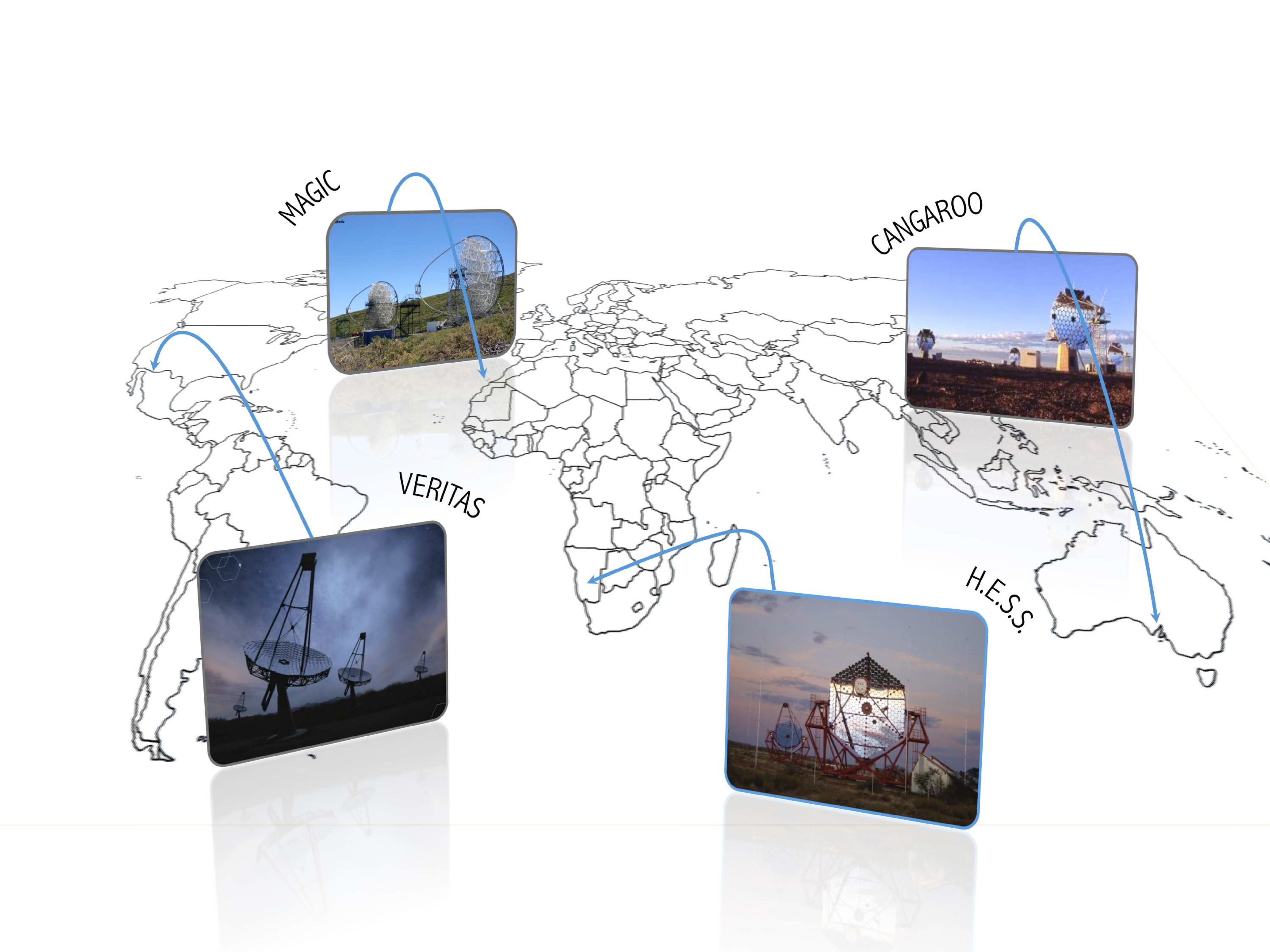}
  \caption{Current generation of Cherenkov telescopes observatories around the globe.\label{globe}}
\end{figure}

Current generation of Cherenkov telescopes all run in stereoscopic mode. In the northern hemisphere, MAGIC uses two large telescopes (17 m diameter) in the Canary islands, and VERITAS runs 4 telescopes of 12 m diameter. VERITAS is the successor of the Whipple telescope and is located in Arizona. In the southern hemisphere, the Cangaroo collaboration runs 4 10 m telescopes in Australia, and HESS consists of 4 telescopes of 12 m diameter in Namibia. Fig.~\ref{globe} shows the locations of the observatories on a map.

The Cherenkov technique is complementary to satellite telescopes, such as Fermi. Fermi works between about 300 MeV and 300 GeV, has a full sky coverage with a large field of view, and a relatively small effective area, of order $1\; \rm m^2$. Whereas Cherenkov telescopes run only at night when the sky is clear, Fermi can record data all around the clock, that compensates for the smaller effective area. The basic features one has to keep in mind in order to understand the pros and cons of using Cherenkov telescopes to search for dark matter are the following:

\begin{itemize}
\item
Large effective areas, of order $10^5 \; \rm m^2$
\item
Relatively small fields of view, of order a few degrees
\item
Angular resolution of about 0.1$^\circ$
\item
Energy Threshold between 50 GeV and 200 GeV
\item
10\% energy resolution
\item
A duty cycle allowing $\sim$1000 h of data taking a year
\end{itemize}

\subsection{The future : CTA}

\begin{figure}
  \includegraphics[height=.3\textheight]{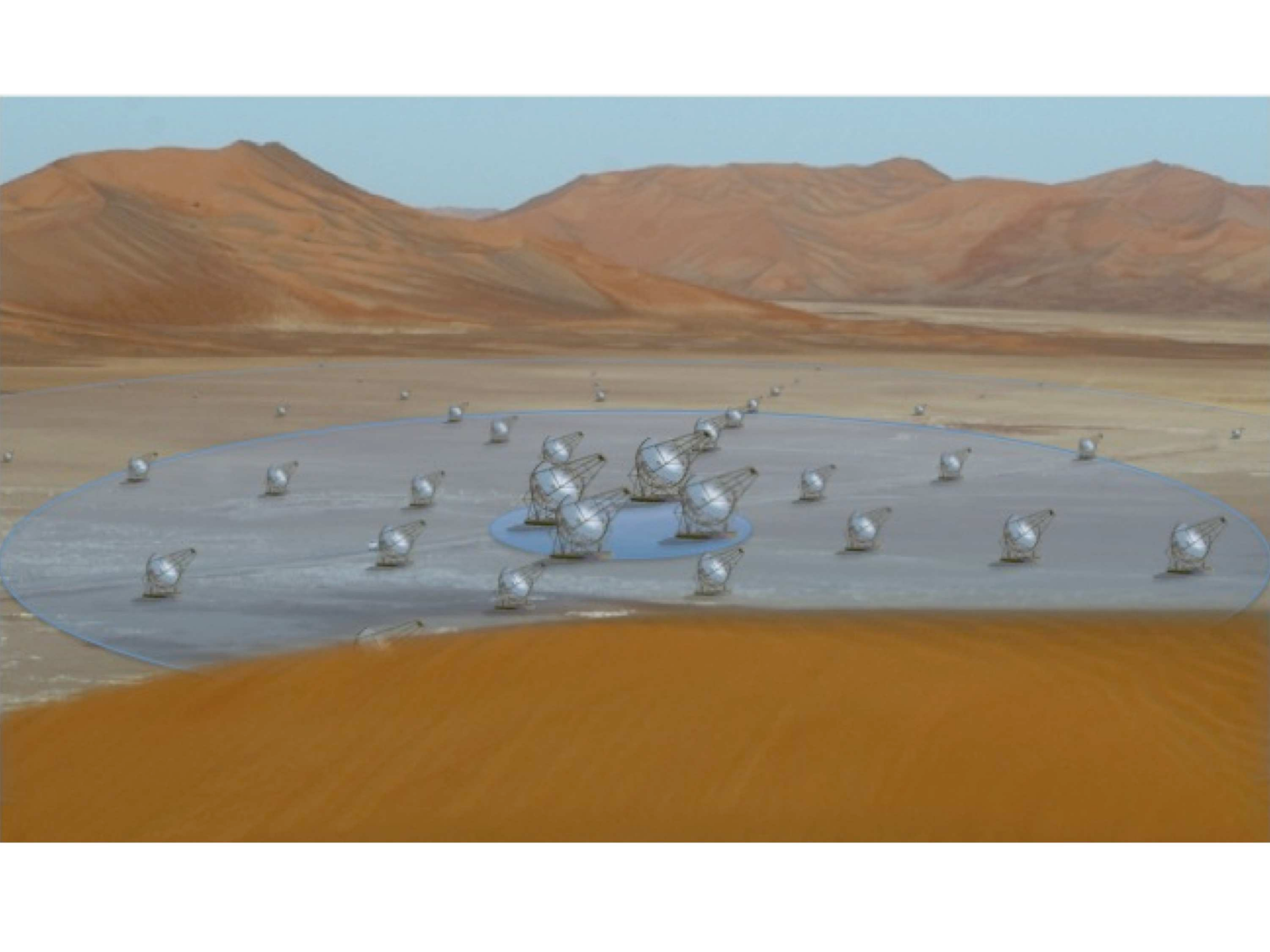}
  \caption{\label{CTA_layout} Artist's view of a possible layout for the CTA observatory: 3 classes of telescopes (23 m, 12 m, 6 m) will allow to cover a wide energy range from 10 GeV to 200 TeV.}
 \end{figure}

The next step for $\gamma$-ray astronomy will consist of the building of a larger array of Cherenkov telescopes --the Cherenkov Telescope Array (CTA)~\cite{Consortium:2010bc}. The observatory will take advantage of proven experimental techniques on HESS, MAGIC and VERITAS, and use 50 to 100 telescopes. One site in each hemisphere will allow a full sky coverage. On each site, the observatory will be composed of 3 groups of telescopes, from the center of the array to its outskirts: large 23 m telescopes, medium 12 m telescopes and small 6 m telescopes, as shown on Fig.~\ref{CTA_layout}. This is intended to cover a wider energy range from a few 10 GeV to 200 TeV. The sensitivity will be improved by a factor of 10 compared to current generation arrays and the angular resolution will be improved by a factor of up to 4. The energy resolution and the duty cycle will be similar to what exist now. The field of view will depend on the class of the telescope, from $\sim$3$^\circ$ for large ones to up to 8$^\circ$ for small ones. This is a technical limitation, as it is difficult to build large telescopes with long focal distance. The layout of CTA will allow a flexible operation: deep observations of a region of the sky involving the whole array, larger surveys with subarrays pointing in different positions on the sky, regular observation with 3/4 of the array while the other quarter monitors variable sources, etc.

\section{Particle dark matter and its indirect detection}

\subsection{Relic density and 2-particles annihilations}

The next half of the paper is devoted to the constraints one can put on particle dark matter models thanks to the good sensitivity of Cherenkov telescopes. This first section recalls the main arguments for particle dark matter and the derivation of possible signals.

The cosmological standard model stipulates that 84\% of the matter in the Universe is non-baryonic. This is motivated by different probes on various scales~\cite{Spergel:2006hy}. As an example, one can notice that the Cosmic Microwave Background (CMB) homogeneity seems to be in conflict with the very existence of galaxies. Indeed, the CMB radiation shows a homogeneous Universe at the level of $\delta = \delta \rho/\rho\sim 10^{-5}$. Nowadays obviously $\delta\gg1$. As $\delta$ grows proportionally to the scale factor of the metric, if the Universe was as homogeneous as the CMB was at recombination, one would have only  $\delta\sim 10^{-2}$ in the present. This simple fact points out the need to introduce a new type of matter which could present larger fluctuations. Not to affect the CMB, this matter should not interact with light, therefore being non-baryonic. In an independent way, models beyond the standard model of particle physics predict the existence of new massive stable particles that have the required properties to make up the cosmological non-baryonic matter, dubbed Dark Matter (DM) in the following. Interestingly enough, these DM candidates were at first not proposed to solve the non-baryonic matter issue but appear when trying to address the limitations of the standard model of particle physics such as the grand unification or the stabilization of the Higgs mass with respect to radiative corrections. The current cosmological DM density is set by their annihilation rate in the early Universe. This provides a natural value for the annihilation cross section of $\langle \sigma v\rangle\sim 3\times 10^{-26}\;\rm cm^3 s^{-1}$, where $\langle \sigma v\rangle$ is the velocity-weighted annihilation cross-section.

\subsection{Expected signals in Cherenkov telescopes and derivation of constraints}

In the standard picture, the stars of our Galaxy lie in a  thin disk of $\sim 20$ kpc radius, which is dipped into a spherical DM halo of scale a factor of $\sim 10$ larger. Some of the DM particles happen to collide within the halo, producing standard model particles. This exotic production of standard model particles is associated with the emission of $\gamma$-ray with energies of the order of the DM particle mass. The considered weakly interactive massive particle (WIMP) being related to electroweak physics, its mass is expected to range between 100 GeV and a few TeV. The $\gamma$-ray emission associated with DM particle annihilations provides a chance to detect DM particles through their collisions.

The expected energy-integrated $\gamma$-ray flux from a region of volume $V$ at a distance $D$ with DM density $\rho$ is 
\begin{equation}
\phi_{\gamma}\;=\;N_{\gamma}\;\frac{\langle \sigma v \rangle}{m^2}\;\frac{1}{4\pi D^2}\;\int_V {\rm d}V \;\frac{\rho^2}{2}\;\;,
\end{equation}
where $N_\gamma$ is the number of $\gamma$-rays per collision, $m$ is the DM particle mass. The number $N_\gamma$ is strongly dependent on the energy threshold of the experiment, as it is the integral of all photons from its threshold to the DM mass. As the expected DM induced photons energy spectra are quickly falling with energy, it is then always better to have a low energy threshold, in order to be sensitive to more photons. The defined volume $V$ can contain a whole DM structure ({\it e.g.} a in dwarf galaxy) or part of it (like for the Milky-Way halo).

Up to now, no convincing DM signal has been observed, so the way to present the results is to draw a contour in the $\sigma v$-$m$ plane. The flux sensitivity of an experiment is determined by the characteristics of the corresponding observation: effective area, exposure time, data quality, zenith angle of the target etc. The experimental work consists precisely in determining the flux sensitivity. Then, for a fixed particle DM mass $m$, a limit on the cross section $\sigma v$ is computed for the expected signal not to be larger than the sensitivity. This can be performed provided the DM density within the observed region is properly modeled. This is the main source of uncertainty in this type of analysis. The constraints are usually compared to model predictions that are computed with codes such as DarkSUSY~\cite{Gondolo:2004sc} or micrOMEGAs~\cite{Belanger:2010gh}. Depending on the particle physics model, the dependence of $N_\gamma$ on the mass of the DM particle is different. Some collaborations (Whipple, VERITAS, sometimes HESS) choose to use a generic parameterization for $N_\gamma(m)$ which is not exact but somehow gives the average value over a possible range of models. It is also possible to extract the value of $N_\gamma$ from the model and compute the constraint point by point. That is the way MAGIC presented its results and in that case the constraint is not a continuous line. Another approach is to show the constraints obtained with the two extreme final sates which are $\tau^+\tau^-$ and $b\bar{b}$.

As for the choice of the targets, the haloes of galaxies and clusters of galaxies are assembled through the merging of a huge number of smaller structures. Most mergers are incomplete and large CDM halos, {\it e.g.} the one around the Milky Way, harbor an enormous population of subhaloes, which are a record of its assembly history.  Some of these subhaloes, for the most massive of them, contain baryonic matter and stars, making up the satellites of the Milky-Way. A careful study of the kinematics of those stars allow obtaining constraints on the DM mass content of the satellite, which can help to reduce the uncertainty on the constraint. However, the presence of baryons in the center of haloes can modify the density profile of the DM. Then, only subhaloes that do not contain baryons are expected to have their primordial shape untouched. The interaction of the satellite or the subhalo with the gravitational potential of the Milky-Way can also cause the profile to be significantly modified by tidal effects. In this review we shall see results from all types of sources: dwarf galaxies with strong tidal disruption or none of it, globular clusters for which baryons could have played a role, the Galactic center where there is an actual conventional $\gamma$-ray source, and DM clumps, which are basically untouched subhaloes. In all cases, the ideal signal one would expect from a dense DM region has the following features:

\begin{itemize}
\item
Slightly extended, with an extension of order 1$^\circ$ and a halo-type morphology
\item
Steadiness: it should have strictly no time variability
\item
No high energy counterpart
\item 
No nearby conventional source
\item
Cutoff at electroweak-like scale in the energy spectrum
\item
A few alter-egos with the exact same properties in the sky
\end{itemize}

\subsection{Searches towards known targets}

The most used strategy is to select targets where the WIMP annihilation process can occur efficiently, and observe them with a sufficient amount of time. Targets are selected thanks to the observation of stars and study of their kinematics. This allows to infer the DM mass content of the target, although with usually large uncertainties. The knowledge of their positions in the sky is very precious for Cherenkov telescopes, because of their relatively small fields of view and background subtraction techniques. The presence of baryons can have drawbacks too, they can alter the DM content of the target or induce conventional $\gamma$-ray emissions. 

\begin{figure}[h]
\centering
\includegraphics[width=\textwidth]{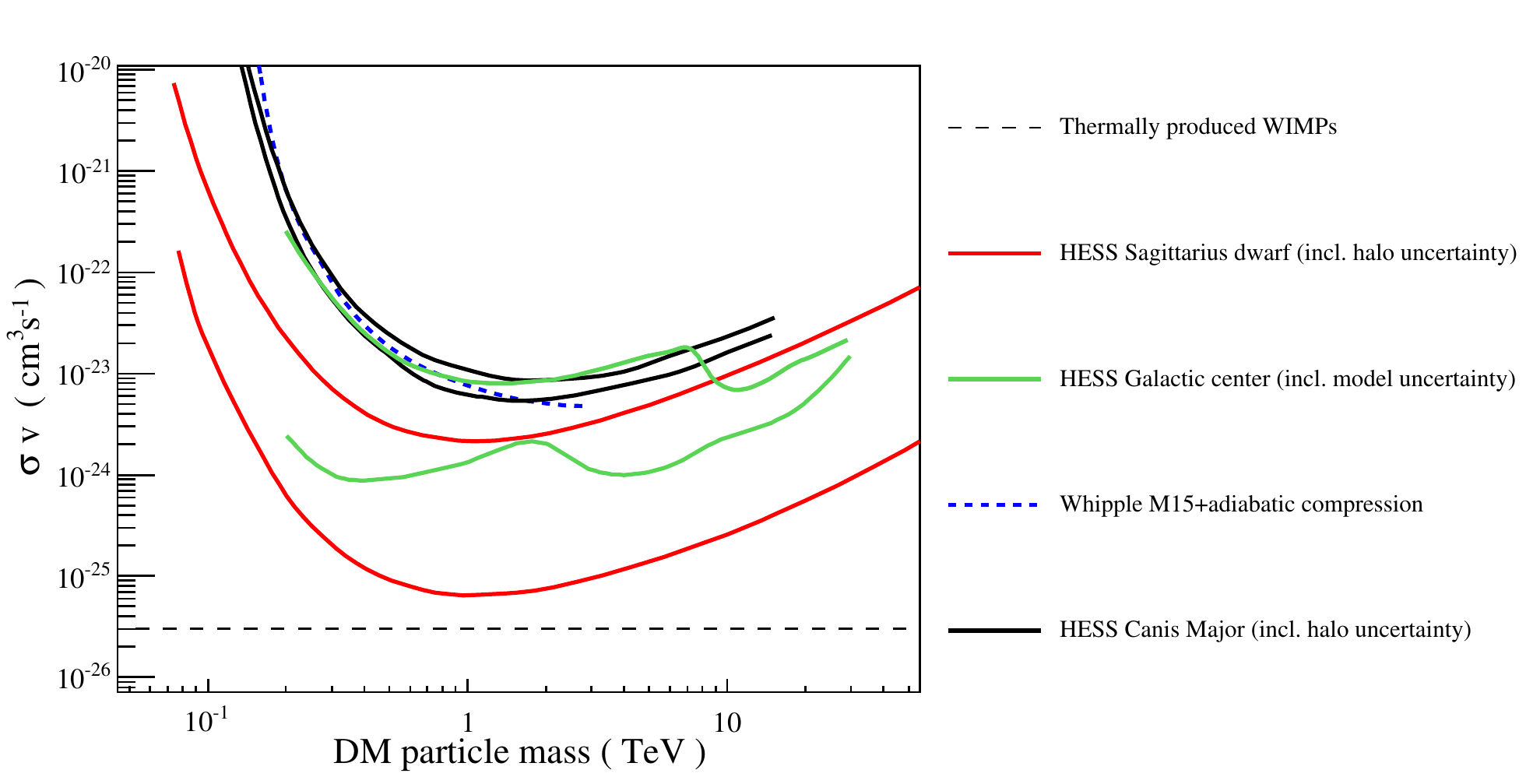}
\caption{Compilation of constraints in the $\sigma v$--m plane, from HESS, Veritas, Whipple, in the case of targets for which more uncertainties are associated (Canis Major overdensity, Galactic center, M15+adiabatic contraction, Sagittarius dwarf galaxy)\label{targets2}}
\end{figure}

The most dense region of our Galaxy is its center, located at 8.5 kpc. HESS dedicated almost 50h of observation to the Galactic center~\cite{Aharonian:2006wh}. Among all the targets discussed here, it is the only one for which a $\gamma$-ray emission has actually been found. It has been shown that most of the emission cannot be due to DM annihilations. The reason is that the energy spectrum of the source does not fit a DM-like spectrum. One should note also that unlike the other targets mentioned here, several astrophysical objects that could be at the origin of the emission are indeed present in the same field of view.
To infer constraints in the $\langle \sigma  v \rangle$-m plane, the authors of~\cite{Aharonian:2006wh} search for the maximal cross section for which the energy spectrum does not exclude the corresponding DM contribution. Results are displayed in Fig.~\ref{targets2}, in which the band represents the uncertainty from the particle physics models on $N_\gamma$.

\begin{figure}
\centering
\includegraphics[width=1.\textwidth]{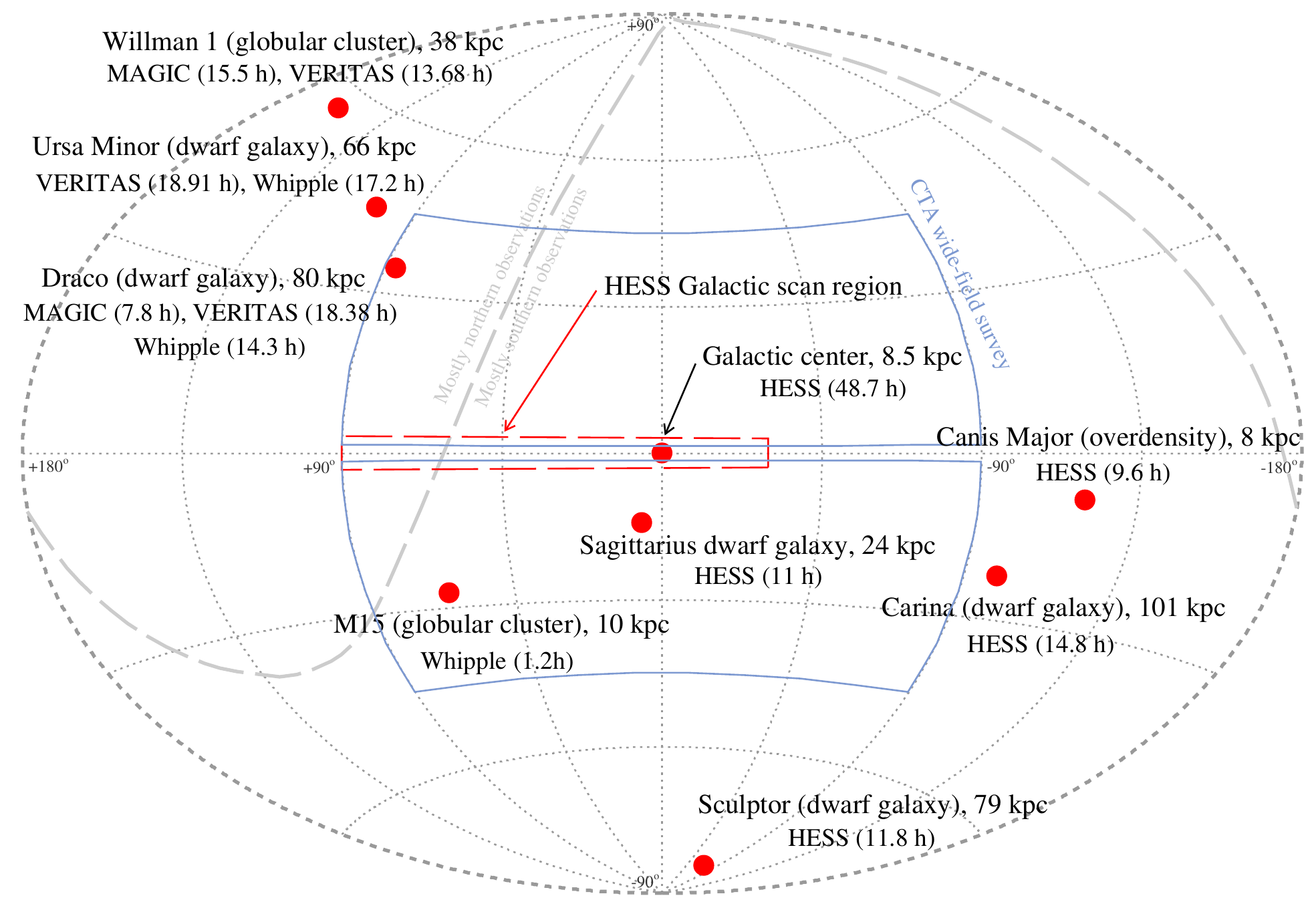}
\caption{Galactic coordinates representation of the targets for which the constraints are displayed in Fig.~\ref{targets} and Fig.~\ref{targets2}, and their main characteristics: distance, type and experiments that observed them, with corresponding exposure. The displayed targets are the Galactic center (HESS~\cite{Aharonian:2006wh}), Sagittarius dwarf (HESS~\cite{Aharonian:2007km}), Sculptor (HESS~\cite{:2010zz}), Carina (HESS~\cite{:2010zz}), Draco (MAGIC~\cite{Albert:2007xg}, VERITAS~\cite{veritas:2010pja}, Whipple~\cite{Wood:2008hx}), Willman I (VERITAS~\cite{veritas:2010pja}, MAGIC~\cite{Aliu:2008ny}), Ursa Minor (VERITAS~\cite{veritas:2010pja}, Whipple~\cite{Wood:2008hx}), Canis Major (HESS~\cite{2009ApJ...691..175A}) and M15 (Whipple~\cite{Wood:2008hx})\label{map_obs}}
\end{figure}

\begin{figure}[h]
\centering
\includegraphics[width=\textwidth]{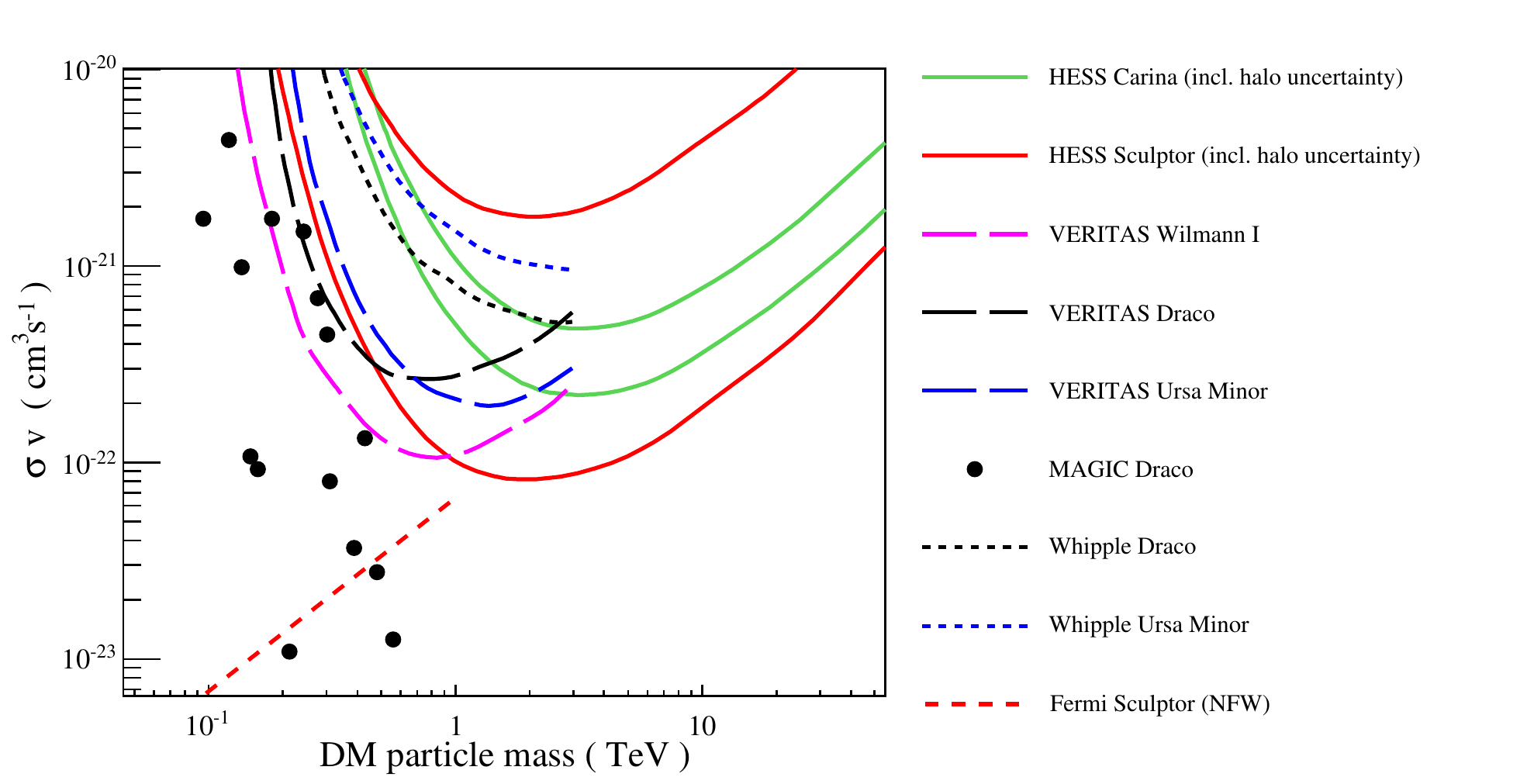}
\caption{Compilation of constraints in the $\sigma v$--m plane, from HESS, Veritas, MAGIC, Whipple, in the case of targets poorly affected by tidal stripping (Carina, Draco, Sculptor, Ursa Minor, Wilmann I).\label{targets}}
\end{figure}

Other observations have been conducted by HESS, Veritas, MAGIC and Whipple. All of them get their constraints from the fact that no signal has been found. Fig.~\ref{map_obs} presents a summary some of the observed targets, with their types, distances, the corresponding experiment and exposure used to set the constraint. Only objects for which constraints are drawn in Fig.~\ref{targets} and Fig.~\ref{targets2} appear in the figure, other objects have been used for searching for particle DM, such as Bo\"{o}tes~\cite{veritas:2010pja}, M33 and M32~\cite{Wood:2008hx} and the Perseus galaxy cluster~\cite{Aleksic:2009ir}, but those are not listed here.
The results from different analyses are shown on Fig.~\ref{targets} and Fig.~\ref{targets2}. They are sort into two groups, depending on wether the DM content modeling suffer from important systematic uncertainties or not. The first group, on Fig.~\ref{targets} concerns targets that are not strongly affected by tidal effects. Fig.~\ref{targets2} displays constrains from somehow more optimistic analyses. It includes observations of the Canis Major overdensity, Sagittarius dwarf galaxy (that experienced strong tidal stripping), the results from the Galactic center, as well as the results from M15, for which a very optimistic assumption has been made (adiabatic contraction around a hypothetical central black hole). As explained in the previous section, the constraints have different shapes because of the parameterization of $N_\gamma$ and the different energy dependence of the effective area of the experiments. For some of the results, an uncertainty band represents the variation of the constraints when the DM content of the target is varied within the region that is compatible with star kinematics (called halo uncertainty in the legends). For the sake of comparison, the result from Fermi on Sculptor dwarf galaxy is shown on Fig.~\ref{targets}. One can note that the constraints are very complementary. The advantage of Fermi here is the much lower energy threshold (below 1 GeV) and the very long exposure. In the case of "safe" dwarves, the limits on $\langle \sigma v \rangle$ lie in the $10^{-21}-10^{-23}\;\rm cm^3s^{-1}$ region. For the more uncertain targets the limits lie between $10^{-23}\;\rm cm^3s^{-1}$ and $10^{-25}\;\rm cm^3s^{-1}$.

The next generation of Cherenkov telescopes, CTA, will certainly conduct the same type of observation. The effective area will be a factor of 10 larger, and constraints will benefit from a lower threshold. In~\cite{Bringmann:2008kj} though, the authors show that only for optimistic particle physics scenarios CTA will be sensitive to natural WIMPs.

\subsection{Blind searches for DM clumps}

\begin{figure}
\centering
\includegraphics[width=.49\textwidth]{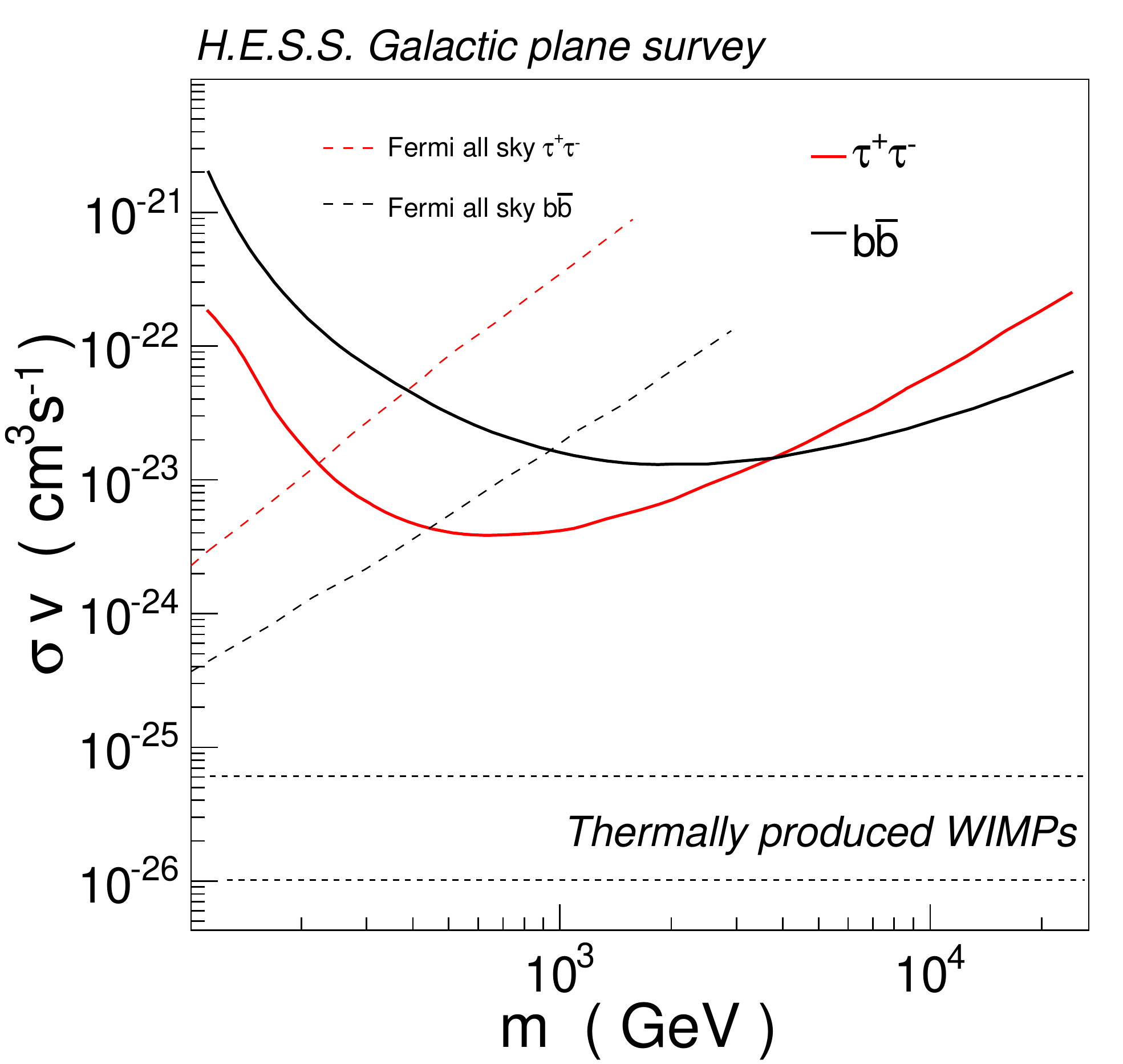}
\includegraphics[width=.49\textwidth]{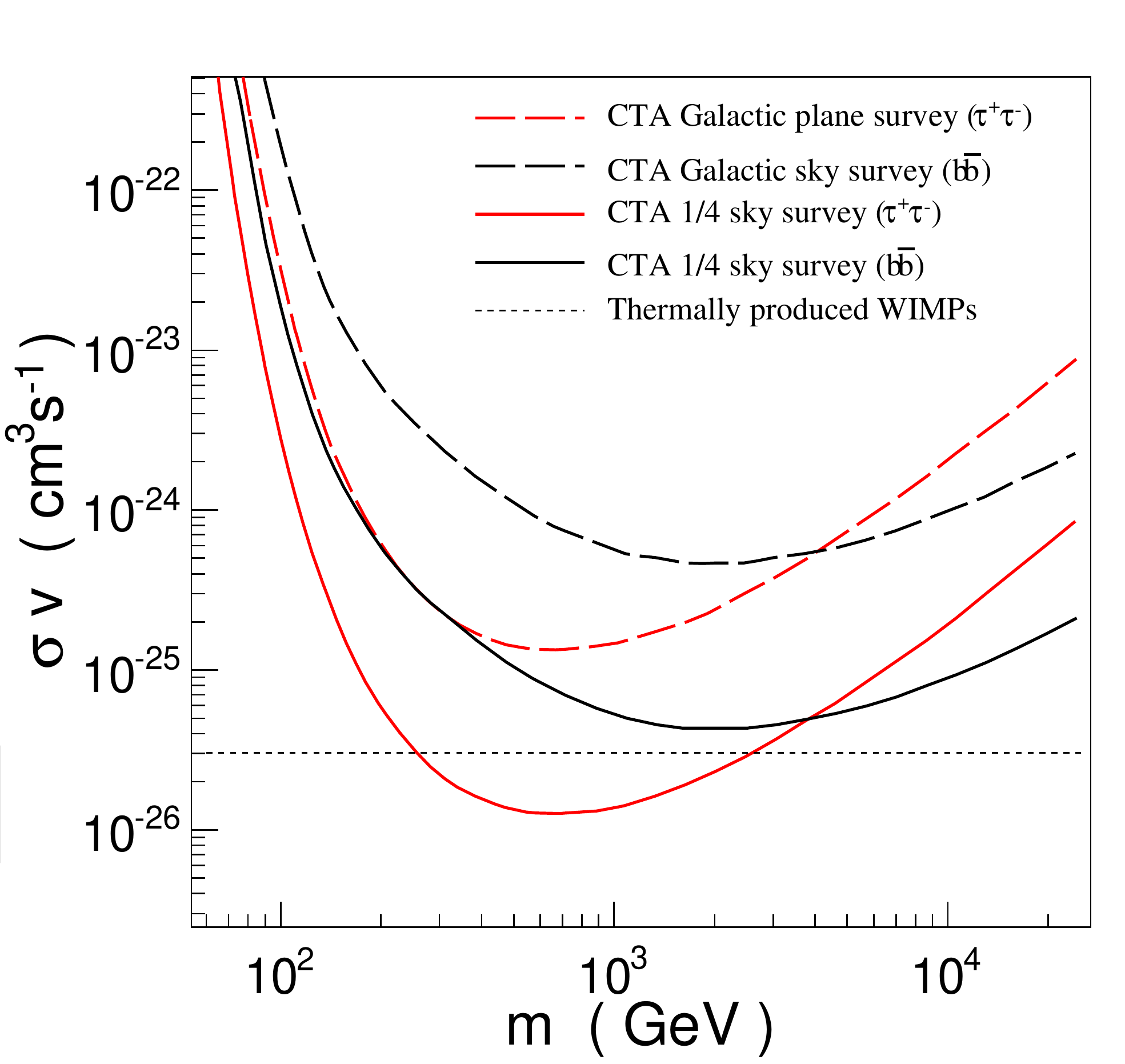}
\caption{Constraints obtained from the search for DM clumps with HESS wide field galactic plane survey (left panel) and prospects for CTA (right panel) in the case of a galactic plane survey and a survey of one fourth of the sky.\label{constraints_survey}}
\end{figure}

Although Cherenkov telescopes cannot perform wide-field surveys in a single shot, the maturity of this technique allows now to conduct scans of significant fractions of the sky, and make blind searches for new sources. This has been done with the HESS experiment, for scanning the Galactic plane~\cite{Aharonian:2005kn}. This type of scan allows searching for DM clumps. Indeed the Milky Way halo is the result of the merging of a large number of smaller haloes, some of which are still present in the Milky Way, as predicted in high resolution N-body simulations~\cite{Diemand:2008in}. In~\cite{Aharonian:2008wt}, data from the HESS survey has been used to build a sensitivity map. The authors of~\cite{Brun:2010ci} have shown that a large number of subhaloes are predicted to be in the scanned region (displayed on Fig.~\ref{map_obs} as a dashed rectangle). For given values of $\sigma v$ and $m$, annihilation within those objects could make a statistically significant number of them shine enough in $\gamma$-rays to be observable. From the observation perspective, none of the unidentified sources discovered in the HESS scan present the required characteristics to be DM clumps. Then, from the convolution of the sensitivity map with the prediction from DM clustering in numerical simulations, it is possible to get the constraints that are shown on the left panel of Fig.~\ref{constraints_survey}. As shown on this figure, the results are again very complementary to Fermi wide field searches~\cite{Buckley:2010vg} although Fermi has a full sky coverage. The drawback from a much smaller scanned regions is compensated in the case of Cherenkov telescopes by a better flux sensitivity. The obtained constraints are of the same order of magnitude as for targeted searches. Note however that the systematic errors on the DM distribution --although still present-- are very different in the two cases. 

Prospects for blind searches for DM subhaloes take advantage of the fact that large surveys will be conducted by CTA independently of the search for DM. At least 2 large scans can be foreseen:  a Galactic plane survey and a wider survey of one fourth of the sky. In~\cite{Brun:2010ci}, constraints on DM models are computed assuming no DM subhalo candidate is found. In the case of a HESS-like survey, constraints are computed with the same exposure. That would allow for an improvement of the HESS constraints by a factor of 10 (see right panel of Fig.~\ref{constraints_survey}). To go further, the 1/4 sky survey will have to be used. In that case, in~\cite{Brun:2010ci} the scan regions has been optimized: centered on the Galactic center, with the exclusion of latitudes between $\pm 1.5^{\circ}$, (blue region in Fig.~\ref{map_obs}). With the same assumptions, and within 6 years of data taking, this allows to reach the thermally produced WIMPs region, as shown on right panel of Fig.~\ref{constraints_survey}.

\section{conclusions}

Cherenkov telescopes now allow for deep searches for particle dark matter both from identified targets and performing blind searches. Those two types of methods have very different types of systematic errors and reach the same level of constraints. CTA will certainly give the opportunity to dig deeper into the models, in particular in the light of the future results from current observations by the Fermi satellite.

\bibliography{brun}

\begin{thebibliography}{22}
\expandafter\ifx\csname natexlab\endcsname\relax\def\natexlab#1{#1}\fi
\expandafter\ifx\csname bibnamefont\endcsname\relax
  \def\bibnamefont#1{#1}\fi
\expandafter\ifx\csname bibfnamefont\endcsname\relax
  \def\bibfnamefont#1{#1}\fi
\expandafter\ifx\csname citenamefont\endcsname\relax
  \def\citenamefont#1{#1}\fi
\expandafter\ifx\csname url\endcsname\relax
  \def\url#1{\texttt{#1}}\fi
\expandafter\ifx\csname urlprefix\endcsname\relax\def\urlprefix{URL }\fi
\providecommand{\bibinfo}[2]{#2}
\providecommand{\eprint}[2][]{\url{#2}}

\bibitem[{\citenamefont{V\"olk and
  Bernl\"ohr}(2009)}]{springerlink:10.1007/s10686-009-9151-z}
\bibinfo{author}{\bibfnamefont{H.}~\bibnamefont{V\"olk}} \bibnamefont{and}
  \bibinfo{author}{\bibfnamefont{K.}~\bibnamefont{Bernl\"ohr}},
  \bibinfo{journal}{Experimental Astronomy} \textbf{\bibinfo{volume}{25}},
  \bibinfo{pages}{173} (\bibinfo{year}{2009}), ISSN \bibinfo{issn}{0922-6435}.

\bibitem[{\citenamefont{Aharonian
  et~al.}(2006{\natexlab{a}})}]{Aharonian:2006pe}
\bibinfo{author}{\bibfnamefont{F.}~\bibnamefont{Aharonian}}
  \bibnamefont{et~al.} (\bibinfo{collaboration}{H.E.S.S.}),
  \bibinfo{journal}{Astron. Astrophys.} \textbf{\bibinfo{volume}{457}},
  \bibinfo{pages}{899} (\bibinfo{year}{2006}{\natexlab{a}}),
  \eprint{astro-ph/0607333}.

\bibitem[{\citenamefont{Vivier}(2009)}]{vivier}
\bibinfo{author}{\bibfnamefont{M.}~\bibnamefont{Vivier}}, \bibinfo{journal}{PhD
  thesis} \textbf{\bibinfo{volume}{IRFU-09-04-T}} (\bibinfo{year}{2009}).

\bibitem[{\citenamefont{CTA-consortium}(2010)}]{Consortium:2010bc}
\bibinfo{author}{\bibnamefont{CTA-consortium}} (\bibinfo{year}{2010}),
  \eprint{1008.3703}.

\bibitem[{\citenamefont{Spergel et~al.}(2007)}]{Spergel:2006hy}
\bibinfo{author}{\bibfnamefont{D.}~\bibnamefont{Spergel}} \bibnamefont{et~al.}
  (\bibinfo{collaboration}{WMAP Collaboration}),
  \bibinfo{journal}{Astrophys.J.Suppl.} \textbf{\bibinfo{volume}{170}},
  \bibinfo{pages}{377} (\bibinfo{year}{2007}), \eprint{astro-ph/0603449}.

\bibitem[{\citenamefont{Gondolo et~al.}(2004)}]{Gondolo:2004sc}
\bibinfo{author}{\bibfnamefont{P.}~\bibnamefont{Gondolo}} \bibnamefont{et~al.},
  \bibinfo{journal}{JCAP} \textbf{\bibinfo{volume}{0407}}, \bibinfo{pages}{008}
  (\bibinfo{year}{2004}), \eprint{astro-ph/0406204}.

\bibitem[{\citenamefont{Belanger et~al.}(2010)}]{Belanger:2010gh}
\bibinfo{author}{\bibfnamefont{G.}~\bibnamefont{Belanger}} \bibnamefont{et~al.}
  (\bibinfo{year}{2010}), \eprint{1004.1092}.

\bibitem[{\citenamefont{Aharonian
  et~al.}(2006{\natexlab{b}})}]{Aharonian:2006wh}
\bibinfo{author}{\bibfnamefont{F.}~\bibnamefont{Aharonian}}
  \bibnamefont{et~al.} (\bibinfo{collaboration}{H.E.S.S.}),
  \bibinfo{journal}{Phys. Rev. Lett.} \textbf{\bibinfo{volume}{97}},
  \bibinfo{pages}{221102} (\bibinfo{year}{2006}{\natexlab{b}}),
  \eprint{astro-ph/0610509}.

\bibitem[{\citenamefont{Aharonian
  et~al.}(2008{\natexlab{a}})}]{Aharonian:2007km}
\bibinfo{author}{\bibfnamefont{F.}~\bibnamefont{Aharonian}}
  \bibnamefont{et~al.} (\bibinfo{collaboration}{H.E.S.S.}),
  \bibinfo{journal}{Astropart. Phys.} \textbf{\bibinfo{volume}{29}},
  \bibinfo{pages}{55} (\bibinfo{year}{2008}{\natexlab{a}}),
  \bibinfo{note}{erratum-ibid.33:274,2010}, \eprint{0711.2369}.

\bibitem[{\citenamefont{Abramowski et~al.}(2010)}]{:2010zz}
\bibinfo{author}{\bibfnamefont{A.}~\bibnamefont{Abramowski}}
  \bibnamefont{et~al.} (\bibinfo{collaboration}{HESS Collaboration})
  (\bibinfo{year}{2010}), \bibinfo{note}{accepted in Astropart. Phys.},
  \eprint{1012.5602}.

\bibitem[{\citenamefont{Albert et~al.}(2008)}]{Albert:2007xg}
\bibinfo{author}{\bibfnamefont{J.}~\bibnamefont{Albert}} \bibnamefont{et~al.}
  (\bibinfo{collaboration}{MAGIC}), \bibinfo{journal}{Astrophys. J.}
  \textbf{\bibinfo{volume}{679}}, \bibinfo{pages}{428} (\bibinfo{year}{2008}),
  \eprint{0711.2574}.

\bibitem[{\citenamefont{Acciari et~al.}(2010)}]{veritas:2010pja}
\bibinfo{author}{\bibfnamefont{A.}~\bibnamefont{Acciari}} \bibnamefont{et~al.}
  (\bibinfo{collaboration}{VERITAS}), \bibinfo{journal}{Astrophys. J.}
  \textbf{\bibinfo{volume}{720}}, \bibinfo{pages}{1174} (\bibinfo{year}{2010}),
  \eprint{1006.5955}.

\bibitem[{\citenamefont{Wood et~al.}(2008)}]{Wood:2008hx}
\bibinfo{author}{\bibfnamefont{M.}~\bibnamefont{Wood}} \bibnamefont{et~al.}
  (\bibinfo{year}{2008}), \eprint{0801.1708}.

\bibitem[{\citenamefont{Aliu et~al.}(2009)}]{Aliu:2008ny}
\bibinfo{author}{\bibfnamefont{E.}~\bibnamefont{Aliu}} \bibnamefont{et~al.}
  (\bibinfo{collaboration}{MAGIC}), \bibinfo{journal}{Astrophys. J.}
  \textbf{\bibinfo{volume}{697}}, \bibinfo{pages}{1299} (\bibinfo{year}{2009}),
  \eprint{0810.3561}.

\bibitem[{\citenamefont{Aharonian et~al.}(2009)}]{2009ApJ...691..175A}
\bibinfo{author}{\bibfnamefont{F.}~\bibnamefont{Aharonian}}
  \bibnamefont{et~al.}, \bibinfo{journal}{Astrophys. J.}
  \textbf{\bibinfo{volume}{691}}, \bibinfo{pages}{175} (\bibinfo{year}{2009}),
  \eprint{0809.3894}.

\bibitem[{\citenamefont{Aleksic et~al.}(2009)}]{Aleksic:2009ir}
\bibinfo{author}{\bibfnamefont{J.}~\bibnamefont{Aleksic}} \bibnamefont{et~al.}
  (\bibinfo{collaboration}{MAGIC}) (\bibinfo{year}{2009}), \eprint{0909.3267}.

\bibitem[{\citenamefont{Bringmann et~al.}(2009)\citenamefont{Bringmann, Doro,
  and Fornasa}}]{Bringmann:2008kj}
\bibinfo{author}{\bibfnamefont{T.}~\bibnamefont{Bringmann}},
  \bibinfo{author}{\bibfnamefont{M.}~\bibnamefont{Doro}}, \bibnamefont{and}
  \bibinfo{author}{\bibfnamefont{M.}~\bibnamefont{Fornasa}},
  \bibinfo{journal}{JCAP} \textbf{\bibinfo{volume}{0901}}, \bibinfo{pages}{016}
  (\bibinfo{year}{2009}), \eprint{0809.2269}.

\bibitem[{\citenamefont{Aharonian
  et~al.}(2006{\natexlab{c}})}]{Aharonian:2005kn}
\bibinfo{author}{\bibfnamefont{F.}~\bibnamefont{Aharonian}}
  \bibnamefont{et~al.} (\bibinfo{collaboration}{HESS Collaboration}),
  \bibinfo{journal}{Astrophys.J.} \textbf{\bibinfo{volume}{636}},
  \bibinfo{pages}{777} (\bibinfo{year}{2006}{\natexlab{c}}),
  \eprint{astro-ph/0510397}.

\bibitem[{\citenamefont{Diemand et~al.}(2008)}]{Diemand:2008in}
\bibinfo{author}{\bibfnamefont{J.}~\bibnamefont{Diemand}} \bibnamefont{et~al.},
  \bibinfo{journal}{Nature} \textbf{\bibinfo{volume}{454}},
  \bibinfo{pages}{735} (\bibinfo{year}{2008}), \eprint{0805.1244}.

\bibitem[{\citenamefont{Aharonian
  et~al.}(2008{\natexlab{b}})}]{Aharonian:2008wt}
\bibinfo{author}{\bibfnamefont{F.}~\bibnamefont{Aharonian}}
  \bibnamefont{et~al.} (\bibinfo{collaboration}{H.E.S.S.}),
  \bibinfo{journal}{Phys. Rev.} \textbf{\bibinfo{volume}{D78}},
  \bibinfo{pages}{072008} (\bibinfo{year}{2008}{\natexlab{b}}),
  \eprint{0806.2981}.

\bibitem[{\citenamefont{Brun et~al.}(2011)\citenamefont{Brun, Moulin, Diemand,
  and Glicenstein}}]{Brun:2010ci}
\bibinfo{author}{\bibfnamefont{P.}~\bibnamefont{Brun}},
  \bibinfo{author}{\bibfnamefont{E.}~\bibnamefont{Moulin}},
  \bibinfo{author}{\bibfnamefont{J.}~\bibnamefont{Diemand}}, \bibnamefont{and}
  \bibinfo{author}{\bibfnamefont{J.-F.} \bibnamefont{Glicenstein}},
  \bibinfo{journal}{Phys. Rev. D} \textbf{\bibinfo{volume}{83}},
  \bibinfo{pages}{015003} (\bibinfo{year}{2011}), \eprint{1012.4766}.

\bibitem[{\citenamefont{Buckley and Hooper}(2010)}]{Buckley:2010vg}
\bibinfo{author}{\bibfnamefont{M.~R.} \bibnamefont{Buckley}} \bibnamefont{and}
  \bibinfo{author}{\bibfnamefont{D.}~\bibnamefont{Hooper}},
  \bibinfo{journal}{Phys. Rev.} \textbf{\bibinfo{volume}{D82}},
  \bibinfo{pages}{063501} (\bibinfo{year}{2010}), \eprint{arXiv:1004.1644}.

\end{thebibliography}

\end{document}